\documentclass[aps,pra,amsfonts,amsmath,amssymb,showpacs]{revtex4}

\input{Qcircuit.tex}

\begin{document}

\title{The Stability of Quantum Concatenated Code Hamiltonians}
\author{Dave Bacon}
\affiliation{Department of Computer Science \& Engineering and Department of Physics, University of Washington, Box 352350,
Seattle, WA 98195 USA} \email{dabacon@cs.washington.edu}
\date{\today}
\begin{abstract}
Protecting quantum information from the detrimental effects of decoherence and lack of precise quantum control is a central
challenge that must be overcome if a large robust quantum computer is to be constructed.  The traditional approach to achieving
this is via active quantum error correction using fault-tolerant techniques.  An alternative to this approach is to engineer
strongly interacting many-body quantum systems that enact the quantum error correction via the natural dynamics of these systems.
Here we present a method for achieving this based on the concept of concatenated quantum error correcting codes.  We define a
class of Hamiltonians whose ground states are concatenated quantum codes and whose energy landscape naturally causes quantum error correction.  We analyze these Hamiltonians for robustness and suggest methods for implementing these highly unnatural Hamiltonians.
\end{abstract}
 \maketitle

\section{Introduction}

It is tempting to consider all physical systems as being information processing devices that manipulate through their dynamics the
information labeled by the system's different states.  This view of computation, however, is severely naive because it leaves out
consideration of whether it is practically possible to robustly store and fault-tolerantly manipulate the information encoded into
the physical system in the presence of noise and imprecise control~\cite{Unruh:95a,Landauer:94a,Landauer:95a}.  For classical
computers, we are blessed with technologies, like the transistor and magnetic media, whose physics guarantees robust classical
computation.  For quantum computers operating according to the quantum circuit model under reasonable error models, the issue of
the existence of robust quantum computation with faulty components was successfully addressed, in theory at least, by the methods
of quantum error correction~\cite{Shor:95a,Steane:96a,Steane:96b} and the subsequent development of the threshold theorem for
fault-tolerant quantum computation~\cite{Shor:96a,Aharonov:97a,Aharonov:99a,Knill:98a,Knill:98b,Kitaev:97b}.  In spite of this
result, the requirements for active quantum error correction are quite severe, and currently no experiment has demonstrated break
even for the most basic active versions of these techniques in a natural setting.  Taking a cue from the classical world, however,
one might wonder whether there exist (or whether we can engineer) quantum physical systems that robustly store information without
the use of active quantum error correction but using only the natural physics of the physical system.  Kitaev was the first to
suggest using the physics of many-body quantum systems, in particular to use a topological order, to provide protection for
quantum information~\cite{Kitaev:97c}.  Following Kitaev's proposal numerous related ideas
~\cite{Kitaev:00a,Barnes:00a,Bacon:01b,Dennis:02a,Freedman:03a,Freedman:03b,Freedman:04a,Freedman:04b,Dorier:05a,Jordan:05a,Weinstein:05a,Bacon:06a}
have been put forward to protect quantum information in a similar manner, some using topological orders and others not topological
in nature.  A main feature of these schemes is that they use quantum error correcting codes along with an energy landscape of a
many-body quantum system to protect quantum information from decoherence.  Roughly one is looking for many-body systems whose
energy levels are quantum error correcting codes and at the same time enforce energetic penalties for quantum errors acting
locally on the system.  None of these methods, however, touch base with the standard method for achieving fault-tolerant quantum
computation which relies on concatenated quantum codes (an exception being the recent work o Vidal and Aguado~\cite{Vidal:08a}.)  Here we present a new method for physically protecting quantum
information based on concatenated coding.

Concatenated codes have a long use in proving the threshold theorem for fault-tolerant quantum
computing~\cite{Aharonov:97a,Aharonov:99a,Knill:98a,Knill:98b,Aliferis:05a}.  The basic idea is to use a hierarchy of encoding into
quantum error correcting codes to protect quantum information.  If encoding quantum information into a quantum error correcting
code can lessen the chance of this quantum information being destroyed by quantum noise, then encoding the encoded quantum
information into a quantum error correcting code should lessen this chance even more.  This is the basic idea of concatenated
quantum error correction.  By concatenating $r$ times one decreases the probability of quantum information being destroyed doubly
exponentially in $r$.  This translates into suppressed of quantum faults at a rate that is  exponential in the number of qubits
used in the concatenated code.  Concatenation is highly reminiscent of renormalization group methods~\cite{Rahn:02a}.

Here we present a class of Hamiltonians, that we call {\em quantum concatenated code Hamiltonians}, whose ground states are
concatenated quantum codes and that have significant protective qualities from decoherence due to the energy landscape of the
Hamiltonian.  The first quality that allows the quantum information to be protected from decoherence is the fact that the ground
state is separated from the first excited state by a constant gap and the smallest quantum error that acts nontrivially on the
degenerate ground state contains error that act on $d^r$ qubits (where $d$ is the distance of the base code and $r$ is the number
of levels of concatenation.)  We will show that this implies that for local perturbations below a certain threshold strength, the
ground state degeneracy is split only exponentially in the number of qubits in our system, i.e. as $e^{-\alpha n}$ where $n$ is
the number of qubits.  We will thus show that for local perturbations below a threshold strength there is a phase of the system
where quantum information encoded into the ground state is protected.  Our detailed analytic derivation of this property is the main contribution of this paper.  This protection is similar to the setting in the toric code
Hamilonian~\cite{Kitaev:97c}.  Further, and beyond the toric code in two spatial dimensions, the energy landscape of our code has
more than just a constant energy barrier between the ground state and the first excited state.  Instead quantum errors that affect
more qubits lead to states of ever increasing energy.  Thus there is an energy barrier that errors that destroy the information
encoded into the ground state must overcome in order to destroy information encoded into the system.  This is similar to the toric
code in four-spatial dimensions~\cite{Dennis:02a}, and provides an example of a self-correcting quantum system~\cite{Bacon:06a}.
Thus, whereas excitations of the toric code Hamiltonian in two-dimensions during computation can disorder information encoded into
the ground state of this system, such excitations on a quantum concatenated code Hamiltonian will not destroy the information.

Quantum concatenated code Hamiltonians in their raw form are not reasonable Hamiltonians for a naturally occuring physical system, since they involve many-qubit interactions on the order of the size of the system.  Thus these Hamiltonians must somehow be engineered into a physical system.  Here we outline one possible approach to achieving this engineering.  This approach involves creating effective Hamiltonians using strong pulsed interactions.  We believe this is one of many possible approaches toward designing systems with quantum concatenated code Hamiltonians, but leave other methods as an open problem for further research.  

In this work we concentrate on the properties of quantum concatenated code Hamiltonians and one very simplistic method for constructing these Hamiltonians using one and two qubit interactions.  In doing this we leave out some vital considerations for the practicality of these methods.  The first, and most obvious, is that we do not consider the spatial locality in this paper.  The second is that our discussion of the robustness of these models is not yet totally complete.  In particular methods for rigorously showing the robustness of these models under a generic set of noise models have not yet been found.  Similarly we do not consider fully
fault-tolerant constructions, since the methods we use to construct quantum concatenated code Hamiltonians are not themselves
robust to introducing errors into the system.  Finally we note that we address only cursorily methods for performing computation
on these systems.  In many ways, then, our results are akin to early results in the field of quantum error correction, before
fault-tolerant methods were discovered: we have a developed a method for protecting quantum information but do not have a full
theory of fault-tolerant quantum concatenated code Hamiltonians.   Quantum concatenated code Hamiltonians, however, are sufficiently powerful in protecting quantum information that we believe they deserve further careful study.

The layout of the paper is as follows.  In Section \ref{sec:concat} we introduce the basic idea of concatenated code Hamiltonians.
We begin this section by briefly reviewing stabilizer quantum error correcting codes and then introduce stabilizer code
Hamiltonians and their brethren, quantum concatenated code Hamiltonians.  In Section \ref{sec:memory} we analyze the robustness of
the ground state of these quantum concatenated code Hamiltonians with respect to a local perturbing interaction.  We show that if
such perturbations are below a certain threshold strength, then these interactions have an exponentially decreasing impact on the
ground state as a function of the system size.  In Section \ref{sec:gadget} we briefly discuss how bang-bang simulation techniques can be used to engineer effective interactions corresponding to quantum concatenated code Hamiltonians.  In Section
\ref{sec:compute} we briefly discuss how quantum computation on quantum concatenated code Hamiltonians can be performed and discuss the self-correcting properties of these Hamiltonians.  Finally we conclude with a list of the many open problems for quantum concatenated code Hamiltonians in Section~\ref{sec:conc}.

\section{Quantum Concatenated Code Hamiltonians} \label{sec:concat}

In this section we will introduce the notion of a concatenated code Hamiltonian.  We begin by reviewing stabilizer quantum error
correcting codes, a particularly nice set of quantum error correcting codes that we will use exclusively in our constructions.  We
then turn to the definition of stabilizer Hamiltonians.  These are Hamiltonians whose ground state is a stabilizer code subspace
and which have a significant energetic gap to the first excited state of the Hamiltonian.  These Hamiltonians have a long history
in quantum computing going back to the work of Kitaev~\cite{Kitaev:97c}.  Our approach is to define a class of stabilizer
Hamiltonians related to concatenated codes.  We introduce these Hamiltonians in this section, deferring until section
\ref{sec:gadget} how to efficiently construct these Hamiltonians using simulation methods.

\subsection{Stabilizer Quantum Error Correcting Codes}

We will work with stabilizer quantum error correcting codes.   We briefly describe these codes here.  A more detailed discussion
of these codes can be found in references~\cite{Nielsen:00a,Gottesman:97a,Poulin:05a}.

Let ${\mathcal P}_1$ denote the Pauli group on a qubit, i.e. the group represented by Pauli matrices and a possible phase $i^k$,
$k \in \{0,1,2,3\}$: ${\mathcal P}_1=\{i^k I, i^k X, i^k Y, i^k Z\}$.  The Pauli group on $n$ qubits, ${\mathcal P}_n$, is just
the group represented by tensor products of Pauli matrices and a phase $i^k$,  $k \in \{0,1,2,3\}$.  If $P=i^k P_1\otimes \cdots
\otimes P_n$ is a Pauli operator, then the weight of this operator is the number of $P_i$ that are not $I$.  On a similar note we
call an operator $k$-local if it can be written as a sum of Pauli operators with weight $k$.  All elements of the Pauli group
square to $\pm I$ and all elements of the Pauli group either commute with each other, $PQ=QP$ or anti-commute with each other
$PQ=-QP$.

Stabilizer codes~\cite{Gottesman:96a} are defined by the use of a particular subgroup of the Pauli group known as a stabilizer
group.  A stabilizer group is a subgroup of the Pauli group that is abelian and that does not contain $-I$.  Since all of the
elements of ${\mathcal S}$ commute we can simultaneously diagonalize these operators.  Let $S_1,\dots,S_{n-k}$ denote a minimal
generating set for a stabilizer subgroup ${\mathcal S}$.  Since the stabilizer subgroup does not contain $-I$, each stabilizer
element squares to $I$, and thus has eigenvalues $\pm 1$.  Simultaneously diagonalizing the $S_i$ allows us to label subspaces of
the Hilbert space of $n$ qubits by the eigenvalues of these operators.  In particular, we can define a basis $|s_1,\dots,
s_{n-k},m\rangle$ where $s_i \in \{ \pm 1\}$ and $m \in \{0,\dots,2^{k}-1\}$, such that $S_i|s_1,\dots,s_{n-k},m\rangle=s_i
|s_1,\dots,s_{n-k},m\rangle$.  In other words for each of the $2^{n-k}$ possible eigenvalues of the stabilizer generators we
obtain a subspace of dimension $2^k$, that we call a stabilizer subspace.  A stabilizer code is usually defined as the $+1$
simultaneous eigenspace of the $S_i$ operators, $S_i |\psi_C\rangle=|\psi_C\rangle$ for $|\psi_C\rangle$ in this code.  Note
however, that we could equally as well have chosen any $\pm 1$ subspace and obtained an equivalent code.

Given a stabilizer ${\mathcal S}$, the normalizer of ${\mathcal S}$ in the Pauli group, ${\mathcal N}$, is the set of elements of
the Pauli group that commute with the stabilizer ${\mathcal S}$.  Note that the normalizer ${\mathcal N}$ contains the stabilizer
${\mathcal S}$.  If we examine the group ${\mathcal N}/{\mathcal S}$, this group acts nontrivially on the different stabilizer
subspaces.  In particular it is isomorphic to a Pauli group on $k$ qubits.  In other words it is possible to choose these
operators in such a way that they act as logical Pauli operators on the $k$ qubits encoded into the stabilizer code.  We will
denote the $i$th encoded Pauli operator on the logical qubits by $\bar{P}_i$ where $P$ is the corresponding Pauli operator.  Note
that there is no canonical chose for this labeling, just as there is no canonical choice for the generators of the stabilizer
group.  We can then label the degeneracy of the $|s_1,\dots,s_{n-k},m\rangle$ basis via logical qubits,
$|s_1,\dots,s_{n-k},b_1,\dots,b_k\rangle$, where the corresponding encoded Pauli operator acts accordingly, $\bar{P}_i
|s_1,\dots,s_{n-k},b_1,\dots,b_k\rangle= |s_1,\dots,s_{n-k}\rangle \otimes |b_1 \rangle \otimes \cdots \otimes P_i |b_i\rangle
\otimes \cdots \otimes |b_k\rangle$.

We are now in a position to describe the error correcting properties of a stabilizer code.  Since the dimension of a stabilizer
with $n-k$ minimal generators is $2^k$, a stabilizer code encodes $k$ qubits into $n$ qubits.  Suppose that $E$ anticommutes with
a stabilizer generator $S_i$.  Then
$S_i(E|s_1,\dots,s_n,b_1,\dots,b_n\rangle)=-ES_i|s_1,\dots,s_n,b_1,\dots,b_n\rangle=-s_iE|s_1,\dots,s_n,b_1,\dots,b_n\rangle$, so
we see that the effect of $E$ is to flip the sign of eigenvalue of the stabilizer generators with which $E$ anticommutes.  This
implies that all such $E$'s are detectable errors: if we start in the all $+1$ simultaneous eigenspace of the stabilizer generator
and if $E$ is applied to this state, then we can measure the eigenvalue of the stabilizer generators and detect that an error has
occurred.  Let $\{E_a\}$ denote a set of Pauli operators that will represent errors.  Then consider the set made up of pairs of
these errors as $\{E_a^\dagger E_b\}$.  If every $E_a^\dagger E_b$ is not in the normalizer minus the stabilizer, i.e. is a
detectable error or in the stabilizer, then we can use the stabilizer code to correct the errors $\{E_a\}$.  We are mostly
interested in codes that can correct errors $\{E_a\}$ made up of all Pauli operators up to a weight $t$.  A code that can correct
$t$ errors must have a distance $2t+1$ where the distance of the code is the smallest weight Pauli operator that acts nontrivially
on the code subspace.   We label a quantum error correcting code on $n$ qubits that encodes $k$ qubits and has a distance $d$ via
the standard notation $[[n,k,d]]$.

Subsystem codes are a generalization of subspace codes in which instead of encoding information into a subspace, one encodes this
information into a subsystem~\cite{Knill:00a,Kempe:01a,Poulin:05a,Kribs:05b,Bacon:06a}.  For our purposes we will only be concerned
with subsystem stabilizer codes.  A simple way to think about subsystem stabilizer codes is as follows.  Take a $[[n,k,d]]$
stabilizer code.  Now assign $r$ of the logical qubits as {\em gauge qubits}.  We can then encode quantum information into the
remaining $k-r$ encoded qubits.  These $k-r$ qubits are a subsystem of the $k$ encoded qubits.  If we stipulate that we do not
care what happens to quantum information encoded into the gauge qubits, then we are encoding into a subsystem code, where we care
not about preserving information encoded into a subspace, but information encoded into a subsystem.  Such codes can always be
turned into subspace codes, but their recovery operations can be significantly simpler than the corresponding subspace codes
~\cite{Bacon:06a,Aliferis:07a}, since one does not care about errors accumulating on the gauge qubits.  We will denote a stabilizer
subsystem code with $r$ guage qubits, encoding $k$ qubits, and have a distance $d$ by the notation $[[n,k,r,d]]$.

We now turn to describing briefly how stabilizer codes are used to obtain thresholds in quantum error correction.  It is this
scheme that we seek to mimic via physical means.   Suppose we have a quantum error correcting stabilizer code ${\mathcal C}$ that
encodes a single qubit into $n$ qubits and has a distance $d$ and hence can correct $t=\lfloor {d-1 \over 2} \rfloor$ quantum
errors.  If single qubit errors occur probabilistically and independently with probability $p$ on this code, then quantum error
correction on this code will fail when greater than $t$ errors occur (assume perfect encoding and decoding for now as we are
simply trying to give the jist of thresholds in error correction.)  This implies that the probability of failure is $p_{\rm fail}
\leq C p^{t+1}$ for some fixed constant $C$. Concatenating a quantum error correcting code with itself is the process of using
encoded codewords to form a new code.  Concatenating a code with itself produces a new code with $n^2$ qubits, a distance $d^2$,
and a probability of failure less than $C(Cp^{t+1})^{t+1}$.  The procedure for correcting an error on a concatenated code is to
correct the errors on the first level of encoding and then to correct the errors that remain on the next level of concatenation.
Concatenating $r$ times, uses $n^r$ qubits and has a probability of failure given by recursing the above failure probability
formula,
\begin{equation}
{p_{\rm fail} \over {p_*}}=\left({p \over p_*} \right)^{(t+1)^r}, \quad {\rm where} \quad p_*={1 \over C}.
\end{equation}
Thus if $p<p_{*}$ (below a threshold) the probability of failing decreases doubly exponentially in the number of levels of
concatenation and exponentially in the number of qubits.  This scaling is the basic principle behind the threshold for
fault-tolerant quantum computing.  We seek to mimic this concatenation method, but now using the energy structure of a
Hamiltonian.

\subsection{Stabilizer Code Hamiltonians}

We now turn the construction of Hamiltonians for many qubit systems that have energy levels that are error correcting code
subspaces and whose energy landscape enforces energetic costs for quantum errors.  We begin our discussion with what we term {\em
stabilizer code Hamiltonians}.

From a stabilizer $[[n,k,d]]$ code we can easily construct a simple Hamiltonian whose ground state is degenerate with this
degeneracy corresponding to the stabilizer code space.  One way to achieve this (dating back to the work described in
~\cite{Kitaev:97c}, see also~\cite{Jordan:05a}) is to make the Hamiltonian a sum of the stabilizer generators $S_i$,
\begin{equation}
H_{\rm stab}=-{J \over 2} \sum_i S_i, \label{eq:hstab}
\end{equation}
or, if we express this in the $|s_1,\dots,s_{n-k},b_1,\dots,b_k\rangle$, basis
\begin{equation}
H_{\rm stab}=- {J \over 2} \sum_{s_1,\dots,s_{n-k} \in \{\pm 1\}} \sum_{b_1,\dots,b_k \in \{0,1\}} \sum_i s_i |s_1,\dots,
s_{n-k},b_1,\dots,b_k\rangle \langle_1,\dots, s_{n-k},b_1,\dots,b_k |.
\end{equation}
The eigenvalues of this Hamiltonian are between $-{J \over 2}(n-1)$ and $-{J \over 2}(n-1)$ in increasing steps of size $J$.
Every energy level is at least $2^k$-fold degenerate and is separated form the nearest energy levels by $J$.   In fact if we label
the energies of the Hamiltonian by an index $i$ from $0$ to $n-1$, such that the energy of the $i$th energy level is $E_i=-{J
\over 2}(n-1)+Ji$, then the degeneracy of the $i$th energy level is $g_i=2 {n-1 \choose i}$.  Note that since the generators of a
stabilizer code are not unique, Eq.~(\ref{eq:hstab}) is not unique.  We will always assume a fixed set of generators and logical
operators have been chosen for a stabilizer code Hamiltonian.  Further note that this is definitely not the only manner to take
stabilizer elements and form a Hamiltonian whose ground state is a stabilizer code.

Stabilizer Hamiltonians have some remarkable properties that we will now describe.  Since we are working with a code of distance
$d$, all Pauli operators with weight less than $d$ act as either identity or $0$ on the ground subspace.  Let $\{E\}$ denote a set
of detectable errors, that is elements of the Pauli group which anti-commute with at least one stabilizer generator.  Then every
such error will take elements of the grounds state of the Hamiltonian to an excited state, since these operators will always flip
the sign of at least one $s_i$ when acting on a $|s_1,\dots,s_{n-k},b_1,\dots,b_k\rangle$ state.  Thus, for example if our code is
a one qubit error correcting code, then the smallest weight Pauli operator that acts as a non-trivial operator on the ground state
of a stabilzier code Hamiltonian is of weight three.  In other words every weight two Pauli operator acting on the ground state of
this Hamiltonian causes the ground state to be excited to a higher energy level.  Since our system has an energetic gap of size
$J$ every single qubit and two qubit operation ``costs'' an energy of at least $J$.  In general if we have a $[[n,k,d]]$ code, any
weight $w<d$ Pauli operator costs an energy at least $J$.

The first stabilizer Hamiltonian where the discussion was explicitly cast in terms of quantum error correction was Kitaev's toric
code Hamiltonian~\cite{Kitaev:97c}.  Kitaev's toric code is a stabilizer code where the generators are four qubit operators acting
on qubits on links of a square lattice with toric boundary conditions.  If this lattice is of size $L^2$, then Kitaev's code is a
$[[2L^2,4,L]]$ code.  Recently Jordan {\em et al}~\cite{Jordan:05a} introduced stabilizer Hamiltonians in order to provide
energetic protection within the context of adiabatic quantum computing.  In particular they considered small $[[4,1,2]]$ error
detecting codes (see also~\cite{Bacon:01a}) as well as the $[[5,1,3]]$ single qubit error correcting code stabilizer Hamiltonian.
In both Kitaev's construction, Jordan {\em et al}'s constructions, and in the similar constructions of coherence-preserve (also
known as supercoherent) qubits~\cite{Bacon:01b,Bacon:01a}, some forms of decoherence on information encoded into the ground state
can be suppressed.  This is because in all of these models, single, double, or multi-qubit errors excite the ground state to an
excited state that is of a greater energy by some constant energy gap $\Delta$.  Then if, for example, the bath is a simple set of
harmonic oscillators at temperature $T \ll \Delta$, these decoherence mechanisms will be exponentially suppressed due to the lack
of occupancy of bath modes with enough temperature to excite the ground state.

As an example consider of a stabilizer code Hamiltonian, consider the $[[5,1,3]]$ quantum error correcting code.  This is the
smallest stabilizer code that can correct a single qubit error~\cite{Bennett:96a,Knill:97a}.  This code is a nondegenerate code,
meaning that every error of weight less than $3$ is a detectable error and is not an element of the stabilizer.  For the
$[[5,1,3]]$ stabilizer code, the stabilizer Hamiltonian can be chosen to be
\begin{equation}
H=-{J \over 2} \left( XZZXI+IXZZX+XIXZZ+ZXIXZ \right). \label{eq:hstab5}
\end{equation}
The ground state of this Hamiltonian is the subspace of all $+1$ eigenvalue eigenstates of the individual terms in this
Hamiltonian.  The logical operators on this stabilizer code may be taken to be $\bar{P}=P^{\otimes 5}$ (although smaller weight
choices are possible.)

Another example of a stabilizer code Hamiltonian, but this time constructed from a subsystem code and not a subspace code, can be
derived by using the code described in~\cite{Bacon:06a}.  This is a $[[9,1,4,3]]$ subsystem code which encodes a single qubit,
uses $4$ gauge qubits, and has distance $3$.  The stabilizer for this code can be generated by
$\{XXXXXXIII,IIIXXXXXXX,ZZIZZIZZI,IZZIZZIZZ\}$ and thus the stabilizer Hamiltonian can be chosen to be
\begin{equation}
H=-{J \over 2} \left(XXXXXXIII+IIIXXXXXXX+ZZIZZIZZI+IZZIZZIZZ\right). \label{eq:hsub9}
\end{equation}
The stabilizer subspaces for this code is $2^5=32$ dimensional, representing $5$ encoded qubits.  However, $4$ of these qubits are
gauge qubits, degrees of freedom which we do not wish to protect from the detrimental effects of decoherence.  A logical $X$
operation on this subsystem code can be chosen to be $XXXIIIIII$ and a logical $Z$ operation on this subsystem code can be chosen
to be $ZIIZIIZII$.  Note that in subsystem codes there is degeneracy beyond the logical qubit degeneracy and which corresponds to
the gauge qubits.


\subsection{Concatenated Stabilizer Code Hamiltonians}

We now move on to describing concatenated stabilizer code Hamiltonians.  The basic idea here will be to construct Hamiltonians
whose ground states are a concatenated stabilizer code.

We begin by describing the first concatenation.  Assume we are working with a $[[n,1,d]]$ stabilizer code.  Take $n^2$ qubits.  On
blocks of size $n$ of these qubits we can implement a stabilizer code as in Eq.~(\ref{eq:hstab}).  Further, using the logical
Pauli operators on each of these $n$ stabilizer codes, we can implement a stabilizer code on the encoded information.

In particular we can define the $r$th concatenated Hamiltonian recursively
\begin{equation}
H_r=\sum_{i=1}^n H_{r-1}^{(i)}+H_{\rm stab}^{\rm encoded},
\end{equation}
where $H_{r-1}^{(i)}$ is the level $r-1$ concatenated stabilizer code acting on the $i$th block of qubits and $H_{\rm stab}^{\rm
encoded}$ is a stabilizer code acting on the encoded quantum information of each of these blocks.  This Hamiltonian will contain
$n^r-1$ terms with Pauli operators as large as weight $n$.  This later fact is the challenge in implementing this Hamiltonian
physically~\cite{Haselgrove:03a,Bullock:08a,vanDerNest:08a}.  For now, however, assume that you could implement this Hamiltonian.
We will return to the question of implementing this Hamiltonian after we analyze properties of this Hamiltonian with respect to
quantum errors.

For concreteness, here is the concatenated Hamiltonian at $r=2$ for the $[[5,1,3]]$ code, where we use the stabilizer Hamiltonian
described in Eq.~(\ref{eq:hstab5}) and the logical operators on this stabilizer code of $\bar{P}_i=P^{\otimes 5}$:
\begin{eqnarray}
-{2 \over J}H&=&\sum_{i=0}^4 X_{5i+1}Z_{5i+2}Z_{5i+3}X_{5i+4}+X_{5i+2}Z_{5i+3}Z_{5i+4}X_{5i+5}+X_{5i+1} X_{5i+3}Z_{5i+4}Z_{5i+5}+Z_{5i+1} X_{5i+2} X_{5i+4} Z_{5i+5} \nonumber \\
&&+X_1 X_2 X_3 X_4 X_5 Z_6 Z_7 Z_8 Z_9 Z_{10}Z_{11} Z_{12}Z_{13}Z_{14}Z_{15} X_{16}X_{17}X_{18}X_{19} X_{20} \nonumber \\
&&+ X_6 X_7 X_8 X_9 X_{10}Z_{11} Z_{12}Z_{13}Z_{14}Z_{15} Z_{16}Z_{17}Z_{18}Z_{19} Z_{20} X_{21}X_{22}X_{23}X_{24}X_{25} \nonumber \\
&&+X_1 X_2 X_3 X_4 X_5 X_{11} X_{12}X_{13}X_{14}X_{15} Z_{16}Z_{17}Z_{18}Z_{19} Z_{20} Z_{21} Z_{22} Z_{23} Z_{24} Z_{25} \nonumber \\
&&+Z_1 Z_2 Z_3 Z_4 Z_5 X_6 X_7 X_8 X_9 X_{10} X_{16}X_{17}X_{18}X_{19} X_{20} Z_{20} Z_{21} Z_{22} Z_{23} Z_{24} Z_{25}.
\end{eqnarray}
We have used the notation that $R_i$ denotes the operator $R$ acting on the $i$th qubit.

\section{Robustness of Quantum Concatenated Code Hamiltonians with Respect to Local Perturbations} \label{sec:memory}

We will now analyze the properties of concatenated code Hamiltonians with respect to serving as a robust quantum memory.  We will
begin by considering the effect of errors on stabilizer Hamiltonians and then use this understanding to analyze the concatenated
code Hamiltonians.  We first discuss these effects on subspace stabilizer code Hamiltonians and then discuss how these
generalize to subsystem stabilizer code Hamiltonians.

\subsection{Single Qubit Errors on Distance Three Subspace Stabilizer Hamiltonians}

Consider the effect of single qubit perturbations on a stabilizer subspace code Hamiltonian (not concatenated) for a $[[n,1,3]]$
code.  To begin with consider a perturbation of the form
\begin{equation}
xV=x \sum_{i=1}^n \sum_{Q \in \{X,Y,Z\}} \lambda_{Q,i} Q_i,
\end{equation}
where $R_i$ denotes the operator $R$ acting non-trivially only on the $i$th qubit and $x$ is our perturbation expansion parameter.  It will be convenient to let $\max_{Q,i} \lambda_{Q,i} =1$ such that $x$ represents the strength of the strongest single qubit term.  Note that this implies that $||V|| \leq 3n$.  We will use the degenerate perturbation theory described in
Appendix~\ref{sec:kato}. This perturbation represents a static single qubit perturbation.  We will delay until later discussion of
more general setups where our qubits are coupled to baths.

We consider $V$ as a perturbation on $H_{\rm stab}$ of Eq.~(\ref{eq:hstab}) and focus on the effect of this perturbation on both
the ground state of $H$ and the excited states of $H_{\rm stab}$.  Recall that these states are made up of stabilizer subspaces,
often with multiple subspaces corresponding to a highly degenerate eigenvalue.  Further we will work under the assumption that
$x||V|| \ll {J \over 2}$ so that the eigenvalues are not shifted by an amount comparable to the gap in our unperturbed
Hamiltonian.

We will begin by deriving an effective Hamiltonian for the ground state.  The ground state of a $[[n,1,3]]$ stabilizer Hamiltonian
is two-fold degenerate with this degeneracy corresponding to the stabilizer code.  Let $\Pi_0$ denote the projector onto this
degenerate ground state.  Recall that the smallest weight operator that does not have a vanishing action on a stabilizer subspace
for a non-degenerate stabilizer code has weight $d$.

Begin by considering the first order perturbation, $G_0^{(0)} V G_0^{(0)}=\Pi_0 V \Pi_0$.  This term vanishes because $V$ is made
up of a sum of single qubit operators and each single qubit operator acts to take a stabilizer subspace to another stabilizer
subspace, since every single qubit error must anticommute with an element of the stabilizer
\begin{equation}
xA_0^{(1)}=xG_0^{(0)} V G_0^{(0)} =0
\end{equation}
(We can ignore the case where single qubit errors are degenerate as this corresponds to qubits that are in fix states unentangled
with the rest of the qubits and can be discarded.)

Next consider the second order perturbation $G_0^{(0)} V G_0^{(1)} V G_0^{(0)} + G_0^{(1)} V G_0^{(0)} V G_0^{(0)} + G_0^{(0)} V
G_0^{(0)} V G_0^{(1)}$.  Only the first of these can be nonzero because, as we have just said, $G_0^{(0)} V G_0^{(0)}=0$.
Explicitly expanding this term we find that it is
\begin{equation}
\Pi_0  \sum_{i=1}^n \sum_{Q \in \{X,Y,Z\}} \lambda_{Q,i} Q_i \sum_{k \neq 0} {\Pi_k \over (E_k-E_0) }  \sum_{j=1}^n \sum_{R \in
\{X,Y,Z\}} \lambda_{R,j}  R_j \Pi_0.
\end{equation}
Since the distance $d$ of the the stabilizer codes is greater than $2$, every weight $2$ Pauli operator has a trivial action on
stabilizer subspaces, acting either as the identity on the code space or as zero on the code space.  Let $\sigma_2$ denote the
number of terms in the cross expansion among operators in the above equation that act as identity.  Note that $\sigma_2$ is at
least $3n$, since every Pauli operator squares to identity, but at most $9n^2$, since there are only $9n^2$ terms.  In particular
this implies that the second order term is proportional to the projector onto the ground state subspace with a constant depending
on the specifics of the code being used,
\begin{equation}
x^2A_0^{(2)}= x^2 b_0 \Pi_0,
\end{equation}
where we can bound $b_0$
\begin{equation}
|b_0| \leq {\sigma_2 \over J},
\end{equation}
using the fact that the stabilizer Hamiltonian has a energies separated by {\em at least} $J$.  Thus we see that the lowest order
term in our perturbation has the effect of shifting the energy levels by an amount bounded by $x^2 b_0$.  Since $x$ is the
strength of the individual qubit couplings, when $x \ll {\sigma_2 \over J}$ the energy shift has much less strength than the
individual qubit couplings.

If we truncate the above perturbation theory expansion at second order we obtain the effective Hamiltonian for the eigenvalues
arising from the ground state of
\begin{equation}
H_0^{\rm eff}=E_0 \Pi_0(x)+ x^2 b_0 \Pi_0 + M_0,
\end{equation}
where $M_0$ is an operator that acts arbitrarily, but whose operator norm is bounded by
\begin{equation}
||M_0||<J \left( { 12 n x \over J} \right)^3.
\end{equation}
Note that this implies that this part of the effective Hamiltonian gives rise to an error on information encoded into the ground
state that is third order in $x$.  It will turn out that there are second order errors that are more important than these errors,
but already one can see that the effect of encoding is to reduce an error of strength $x$ to a level shift of strength $x^2$ and
an error of strength $x^3$.

Next consider the effect of $xV$ on the excited states of the stabilizer Hamiltonian.  Recall that a given energy level for a
stabilizer Hamiltonian consists of multiple stabilizer subspaces.  Let $i$ denote the $i$th energy level of the stabilizer
Hamiltonian and let $\alpha$ label the different stabilizer subspaces that make up this $i$th energy level.  The first order
perturbation term is given by $G_i^{(0)} V G_i^{(0)}$.  This is no longer zero, since a single qubit error can move one between
the different stabilizer subspaces.  Define $Q_{i,\alpha,\beta}$ as an operator with operator norm less than $1$ that acts between
the $\alpha$th and $\beta$th stabilizer subspaces of the $i$th energy level.  Then
\begin{equation}
x A_i^{(1)}=xG_i^{(0)} V G_i^{(0)}=x \sum_{\alpha \neq \beta} a_i(\alpha,\beta)Q_{i,\alpha,\beta}.
\end{equation}
We can bound $a_1(\alpha,\beta)$ by counting the number of ways that the single qubit errors can take the $\alpha$th stabilizer
subspace to the $\beta$th stabilizer subspace.  Note that this error acts entirely within the excited state energy manifold and
does not connect to any other energy levels.  It is this property that will render this term largely irrelevant to our analysis.

The second order perturbation term is given by $G_i^{(0)} V G_i^{(1)} V G_i^{(0)} + G_i^{(1)} V G_i^{(0)} V G_i^{(0)} + G_i^{(0)}
V G_i^{(0)} V G_i^{(1)}$.  The first of these terms will give rise to a term that acts entirely within the $i$th energy manifold.
Like the first order term it can expressed entirely in terms of operators that act between the stabilizer subspaces,
\begin{equation}
G_i^{(0)} V G_i^{(1)} V G_i^{(0)}= \sum_{\alpha,\beta} b_i(\alpha,\beta) R_{i,\alpha,\beta},
\end{equation}
where $R_{i,\alpha,\beta}$ is an operator whose norm is bounded by $1$ and we can bound $b_i(\alpha,\beta)$ by the number of ways
that two single qubit errors (perhaps operating on the same qubit) take the $\alpha$th stabilizer subspace to the $\beta$th
stabilizer subspace divided by the energy gap that is at least $J$ (coming from the $G_i^{(1)}$ term.)

Next consider the second and third terms in the second order perturbation term.  This is the first term that can take an excited
energy level and move it to another energy level.  Indeed we see that this term can be nonzero when a single qubit error takes the
$\alpha$th stabilizer subspace to a $\beta$th stabilizer subspace of the same energy, and then another single qubit error takes
that $\beta$th stabilizer subspace to another stabilizer subspace of energy different that $i$.  The net effect of this is an
operator that takes the $\alpha$th stabilizer subspace of the $i$th energy level to the $\alpha^\prime$th stabilizer subspace of
the $j$th energy level.  Let $T_{i,j,\alpha,\alpha^\prime}$ denote an operator of maximum norm $1$ that acts between the
$\alpha$th stabilizer subspace of the $i$th energy level and the $\alpha^\prime$th stabilizer subspace of the $j\neq i$th energy
level.  Then the perturbative terms can be written as
\begin{equation}
G_i^{(1)} V G_i^{(0)} V G_i^{(0)} + G_i^{(0)} V G_i^{(0)} V G_i^{(1)}=\sum_{i \neq j} \sum_{\alpha,\alpha^\prime}
b_{ij}(\alpha,\alpha^\prime) T_{i,j,\alpha,\alpha^\prime}.
\end{equation}
Note that we can bound $b_{ij}(\alpha,\alpha^\prime)$ by the number of three qubit processes that cause errors of the form
described above divided by the energy gap between the $i$th and $j$th energy level, which is at least $J$.  Note that this term
can take the ground state to an excited state of the stabilizer Hamiltonian.  Importantly, however, this term is of strength on
the order of $x^2$.

Putting this together we have derived an effective Hamiltonian for the $i$ energy level given by
\begin{equation}
H_i^{\rm eff}= E_i \Pi_i(x)  -x\sum_{\alpha \neq \beta} a_i(\alpha,\beta)Q_{i,\alpha,\beta} +x^2 \sum_{\alpha,\beta}
b_i(\alpha,\beta) R_{i,\alpha,\beta}+x^2\sum_{i \neq j} \sum_{\alpha,\alpha^\prime} b_{ij}(\alpha,\alpha^\prime)
T_{i,j,\alpha,\alpha^\prime}+M_i,
\end{equation}
where again the error term is bounded by
\begin{equation}
||M_i||<J \left( { 12 n x \over J} \right)^3.
\end{equation}

Recall that the effective Hamiltonian $H_i^{\rm eff}$ is exactly the Hamiltonian whose eigenvectors are the eigenvalues arising
from the slitting of the $i$th energy level.  Thus summing over all of the effective Hamiltonians will give us exactly $H=H_0 +x
V$.  In particular we have derived
\begin{equation}
H=H_{\rm stab}^\prime+x^2 b_0 \Pi_0+\sum_{i \neq 0} \left[ x\sum_{\alpha \neq \beta} a_i(\alpha,\beta)Q_{i,\alpha,\beta}+ x^2
\sum_{\alpha,\beta} b_i(\alpha,\beta) R_{i,\alpha,\beta}+x^2\sum_{i \neq j} \sum_{\alpha,\alpha^\prime}
b_{ij}(\alpha,\alpha^\prime) T_{i,j,\alpha,\alpha^\prime} \right]+M, \label{eq:mainham}
\end{equation}
where
\begin{equation}
||M||<J \left( { 12 n x \over J} \right)^3,
\end{equation}
using the fact that each effective Hamiltonian gives rise to perpendicular subspaces and defining
\begin{equation}
H_{\rm stab}^\prime=\sum_{i=0}^{n-1} E_i \Pi_i(x).
\end{equation}
Note that $H_{\rm stab}^\prime$ acts on each new eigenspace in a trivial manner, i.e. only shifts the energy of this energy
eigenspace.  This is not to say that this operator preserves information encoded into the ground state of the original $H_{\rm
stab}$, but rather that the ground state is changed to a different ground state.  In other words, $H_{\rm stab}^\prime$ is not
responsible for splitting the energy levels.  We analyze this term later and show that the new ground state is close enough to the
original ground state, where close enough will be defined in terms of a quantum error correcting code order
parameter~\cite{Bacon:06a}.

Let us focus on how the $H$ we have derived acts on the ground state.  The ground state is acted upon by the first two terms and
last two terms in Eq.~(\ref{eq:mainham}).  For now we will ignore the effects of the first term and focus on the other three
terms.  If our stabilizer code has the basis $|s_1,\dots,s_{n-1},b\rangle$, then we can express a generic operator on our space as
\begin{equation}
\sum_{L \in \{I,X,Y,Z\}} W_L \otimes L,
\end{equation}
where $W_L$ acts only on the stabilizer tensor component of our states,
$W_L|s_1,\dots,s_{n-1},b\rangle=(W_L|s_1,\dots,s_{n-1}\rangle) \otimes|b\rangle$.  In particular we can do this for the second
term and the last two terms in Eq.~(\ref{eq:mainham}),
\begin{equation}
\sum_{L \in \{I,X,Y,Z\}} V_L \otimes L= x^2 b_0 \Pi_0 +x^2\sum_{i \neq j} \sum_{\alpha,\alpha^\prime} b_{ij}(\alpha,\alpha^\prime)
T_{i,j,\alpha,\alpha^\prime} +M.
\end{equation}
When we do this we can further obtain a bound on the strength of $V_L$:
\begin{equation}
||V_L|| \leq x^2 |b_0| + x^2 \max_{i \neq 0} \max_{\alpha} |b_{0i}(0,\alpha^\prime)| + J \left( { 12 n x \over J} \right)^3.
\end{equation}
Note that $|b_{0i}|$ is at most $(3n)^2 \over J$ by a naive counting of all possible processes that can act as two qubit errors of
the correct form. Better bounds can be obtained for specific codes using the combinatorics of how errors occur on such codes.  We
perform such an analysis for the five qubit code in Section~\ref{sec:5}.

It is useful to take a step back and examine what we have shown so far.  We have shown that it is possible to describe the effect
of a single qubit perturbation on a stabilizer Hamiltonian in a form that turns the stabilizer Hamiltonian into a new perturbed
stabilizer Hamiltonian $H_{\rm stab}^\prime$ and acts on the ground state only nontrivially in order $x^2$, i.e. of strength less
than $C x^2$ for some constant $C$ and some small enough $x$.  Thus if $x<{1 \over C}$ this interaction strength is less than the
single qubit coupling strengths and we have effectively reduced the strength of interactions that can cause information encoded
into the ground state to be destroyed.  Thus we see that by encoding information into the ground state of this new Hamiltonian,
single qubit perturbations can destroy this information, but do so to a lesser degree than if we had not encoded the information.
But what about this new perturbed Hamiltonian, $H_{\rm stab}^\prime$?

We will now show the eigenspaces of $H_{\rm stab}^\prime$ are actually very close to those of $H_{\rm stab}$, in a well defined
manner.  In particular we will calculate the projector onto the $i$th perturbed subspace, $\Pi_i(x)$ via the perturbation theory
describe in Appendix \ref{sec:kato}.  To first order, we find that
\begin{equation}
\Pi_i(x)=\Pi_i - x(G_i^{(1)} V G_i^{(0)} + G_i^{(0)} V G_i^{(1)}) + N_i,
\end{equation}
where we can bound
\begin{equation}
||N_i|| \leq 2 \left({ 12 n x  \over J} \right)^2.
\end{equation}
Note that the first order term does not vanish.  This first (and higher) order terms appearing here imply that the eigenstates of
$H_{\rm stab}^\prime$ are no longer made up of stabilizer subspaces.  What do these eigenstates look like?  Suppose we take a
stabilizer subspace state $|\psi_i\rangle$ and apply the above projector.  Then we will obtain
\begin{equation}
\Pi_i(x)|\psi_i\rangle = |\psi_i\rangle- x (G_i^{(1)} V G_i^{(0)} + G_i^{(0)} V G_i^{(1)})|\psi_i\rangle + x^2 \alpha
|\phi_i\rangle,
\end{equation}
where we can bound $\alpha \leq 2\left({ 12 n x  \over J} \right)^2$ and $|\phi_i\rangle$ is some arbitrary state dependent on
$N_i$.  Now notice that $(G_i^{(1)} V G_i^{(0)} + G_i^{(0)} V G_i^{(1)})|\psi_i\rangle$ consists of single qubit errors acting on
the the stabilizer subspace state $|\psi_i\rangle$.  Since our code can correct a single qubit error, this component of the wave
function represents a {\em correctable} error.  Further we can bound the amplitude of this term by noting that every single qubit
error can contribute at most ${x \over J}$ amplitude to this state.   In other words, the new energy levels of $H_{\rm
stab}^\prime$ are made up of a superposition over the original stabilizer code state plus single qubit error terms acting on these
states of relative amplitude $O(x)$, plus a term whose relative amplitude is $O(x^2)$.

At this point it is important to introduce a quantum error correcting code order parameter~\cite{Aliferis:05a,Bacon:06a}.  Suppose
that you have information encoded into an error correcting code and then some noise process occurs on this encoded information.
For noise processes of a correctable form, i.e. for say single qubit errors, the information encoded into our code has not really
been destroyed, since there is a measurement we can perform that will allow us to restore the quantum information.  Thus, given a
quantum error correcting code, we can define a quantum error correcting order parameter for stabilizer codes as follows.  Suppose
we are trying to define a quantum error correcting order parameter for information encoded into a stabilizer subspace
$|s_1,\dots,s_{n-1},b\rangle$ for a fixed set of $s_i$.  Correctable errors on this code either act trivially on these states or
as detectable errors with particular syndrome measurements.  In particular for each Pauli error that is correctable and not
trivial, $E$, we can define a syndrome $s_1(E),\dots,s_{n-1}(E)$ which, if we measured these outcomes for the stabilizer
generators, would indicate that error $E$ has occurred.  Then the quantum error correcting order parameter is given by
\begin{equation}
D=P(s_1,\dots,s_{n-1}) +\sum_E E P(s_1(E),\dots,E(s_{n-1})),
\end{equation}
where the sum is over all correctable errors.  The importance of this error correcting order parameter is as follows.  Suppose
that we encode information into $|\phi\rangle$ and then an error process occurs that turns this state into a new state $\rho$.
Then $\langle \phi|D^\dagger \rho D|\phi\rangle$ has contributions from $\rho$ that are $1$ if the information is correctable, and is $0$ otherwise.

Consider what this all means for the use of a stabilizer code Hamiltonian as a quantum memory.  In the presence of a constant and
small perturbing interaction of the form we have considered, two effects occur.  One is that the order $x$ couplings are changed
into order $x^2$ couplings.  The second effect is that the ground state of the Hamiltonian is slightly perturbed.  This
perturbation produces effects to first order that are correctable, but that to second order are not correctable.  Thus both
detrimental effects have changed first order perturbations to second order perturbations.

\subsubsection{Coupling to a Bath}

Finally it is important to note here that while our derivation was for a perturbation from a constant single qubit perturbation,
our results are actually more generic than this.  In particular consider the case where our qubits are coupled to a bath.  Then
our perturbation can be expressed as
\begin{equation}
xV = x \sum_{i=1}^n \sum_{Q \in \{X,Y,Z\}} \lambda_{Q,i} Q_i \otimes B_{Q,i} + H_{\rm bath},
\end{equation}
where $B_{Q,i}$ and $H_{\rm bath}$ act on the bath's Hilbert space.  Then if we absorb in $x$ the strength of the bath operators
into $B_{Q,i}$, such that $||B_{Q,i}|| \leq 1$ and work in the rotating frame of the bath Hamiltonian, then our above derivation
will carry through.

Note, however, an important caveat about this procedure.  By assuming that we have absorbed the strength of $B_{Q,i}$ into $x$ we
have limited the models for coupling of a quantum system to its environment.  In particular if the bath operators are not bounded,
as for example happens when the bath is a harmonic oscillator, then our procedure will not work.  This is a restriction in our
analysis.  We expect, however that it practice it will be less of a problem, since in practice our baths do not have initial
states that have unbounded energy and that will cause the unboundedness to become a problem.  However, a detailed analysis of this
problem is beyond our initial investigation of quantum concatenated code Hamiltonians.

\subsubsection{Example for the Five Qubit Code} \label{sec:5}

Here we derive the combinatorial factors for the five qubit code~\cite{Bennett:96a,Knill:97a} that allow us to explicitly bounded
the effective logical qubit strength.  Recall that Eq.~(\ref{eq:hstab5}) gives a stabilizer Hamiltonian for the $[[5,1,3]]$
subspace code.  $b_0$ is the number of ways that a two qubit error process can produce identity on the stabilizer code.  Since the
five qubit code is a perfect code, the only way that this can happen is when the same two qubit error occurs on a qubit.  This
implies that $|b_0|<{15 \over J}$.

Similarly we can bound $b_{0i}(\alpha)$.  Recall that the energy levels of a stabilizer Hamiltonian are between $-{J \over 2}
(n-1)$ and ${J \over 2}(n-1)$ in steps of $n$, or $E_i=-{J \over 2}(n-1) + Ji$.  The number of stabilizer subspaces that make up
the $i$th energy level is given by ${n-1 \choose i}$.  By explicit computation we can see that the largest error process for the
relevant two qubit errors are ones that cause excitations to the first excited state.  In particular we can bound
\begin{equation}
\left \| \max_{i\neq 0} \max_{\alpha^\prime} b_{0i}(0,\alpha^\prime) \right \| \leq {12 \over J},
\end{equation}
since there are $4$ stabilizer subspaces to excite to and $3$ remaining for the second transition in this second order
perturbation effect.

Putting this together we find that the effective strength of our single qubit interactions on the encoded information has become
\begin{equation}
||V_L|| \leq {15 \over J} x^2  + x^2 {12 \over J} + J \left( { 60 x \over J} \right)^3.
\end{equation}
If we are interested in where this becomes less than $x$, we obtain a polynomial inequality in terms of $\gamma={x \over J}$
\begin{equation}
27 \gamma + 60^3 \gamma^2 \leq 1,
\end{equation}
which is true for $\gamma<{\sqrt{96081}-9 \over 14400} \approx 0.0020901$.

\subsection{Single Qubit Errors on Distance Three Subsystem Stabilizer Hamiltonians}

Above we have consider the effect of single qubit perturbations on a subspace stabilizer Hamiltonian.  But what if we use a
subsystem stabilizer Hamiltonian instead?  In this case, errors which act on the gauge qubits do not have any effect on
information encoded into the logical subsystem.  Much of the analysis which we have performed above carries over to subsystem
stabilizer code Hamiltonians.  However there are a few modifications which we now mention.

First note that because a subspace of the stabilizer Hamiltonian contains a degeneracy corresponding to the logical qubit and in
addition gauge qubits, there can be weight $2$ errors acting on the ground state which act non-trivially on the degeneracy of the
ground state.  Thus, whereas before this gave rise to a $x^2 b_0 \Pi_0$ term, this can now give rise to an $O(x^2)$ term which
acts trivially on the information in the logical qubit, but which acts nontrivially on the gauge qubit degrees of freedom.
This means that instead of giving rise to just a projector onto the ground state subspace of order $x^2$, the gauge qubits may be
split with a strength of $O(x^2)$.  Note however, that such interactions, by definition of the gauge qubits, act trivially on the
information in the logical qubits.  For this reason these errors do not give rise to new problematic interactions for subsystem
codes.  Similarly there are now errors which take an excited energy level to the ground state, but which do not act on the
information encoded into the logical qubits.  It is only the errors on the logical qubits which matter in deriving the effect of
the single qubit perturbations on the encoded information.

\subsubsection{Example for the Nine Qubit Subsystem Code}

For the nine qubit subsystem $[[9,1,4,3]]$ code whose stabilizer Hamiltonian is given in Eq.~(\ref{eq:hsub9}), we can carry out an
analysis of the effect of single qubit perturbations similar to that which we have performed for the five qubit stabilizer
Hamiltonian.  We begin by bounding $b_0$.  Here, there are two qubit errors which act non-trivially on the guage qubits.  However
these interactions only act trivially on the information encoded into the logical subsystem of this code.  Thus we can bound $b_0$
in a way similar to the five qubit code: the only interactions which produce an energy shift come from the same single qubit
interaction acting twice on the subsystem.  This implies that $|b_0|< {27 \over J}$.

We can bound $b_{0i}(\alpha)$ in a matter similar to how we proceeded with the five qubit subspace code.  Again the errors which
are of greatest weight are those which cause the ground state to be excited out of the ground state to the first excited subspace.
By direct enumeration there are $44$ such two qubit errors which contribute to this term.  Thus
\begin{equation}
\left \| \max_{i\neq 0} \max_{\alpha^\prime} b_{0i}(0,\alpha^\prime) \right \| \leq {44 \over J}.
\end{equation}
As before we can put this together to bound the effective interaction strength of the the single qubit interactions on the encoded
quantum information
\begin{equation}
||V_L|| \leq {27 \over J} x^2 +{44 \over J} x^2+J\left({60 x \over J}\right)^3.
\end{equation}
This becomes less that the original single qubit strength $x$, when
\begin{equation}
71 \gamma + 60^3 \gamma^2 < 1,
\end{equation}
where $\gamma={x \over J}$.  This is true for $\gamma< {-71+\sqrt{869041} \over 432000} \approx 0.0019936$.

\subsection{Single Qubit Errors on Concatenated Distance Three Stabilizer Code Hamiltonians}

We now proceed to analyze the effects of errors on a concatenated stabilizer code Hamiltonian.  We will begin by considering again the simplest case where the code is a $[[n,1,3]]$ code.

We have shown above that the effect of a single qubit interaction on a stabilizer code of distance $d=3$ on the ground state is an
operator that acts as a logical operator tensored with an operator that acts to excite the ground state to a higher energy level.
Consider now a concatenated code that has been concatenated only one time.  For each of the $n$ lowest level stabilizer
Hamiltonians, we can apply Eq.~(\ref{eq:mainham}).  If we now focus on the ground state of each of these $n$ Hamiltonians (which
is $2^n$-fold degenerate) we see that the effect of our single qubit perturbations are now turned into single qubit perturbations
on the $n$ encoded qubits tensored with operators that act on the stabilizer space of each relevant encoded qubit along with the
new stabilizer Hamiltonians that have perturbed ground states.  But, as we noted when we discussed our system coupled to a bath
for a single stabilizer code Hamiltonian, we can still apply our effective Hamiltonian procedure, but now to the encoded qubits.
In other words, if the effect of the stabilizer code Hamiltonian is to take single qubit interactions of strength $x$ to strength
$Cx^2$ on encoded qubits, then the effect of these interactions on a concatenated code Hamiltonian for one level of concatenation
will act as encoded single qubit interactions of strength $C(Cx^2)^2$.

We now see that our argument allows us to show that single qubit interactions of strength $x$, at the $r$th level of concatenation
produce an error on the encoded qubits of strength less than
\begin{equation}
x_{r}\leq {1 \over C}(Cx)^{2^r} =x_* \left( x \over x^* \right)^{2^r},
\end{equation}
where $x_*={1 \over C}$ is the {\em threshold coupling strength}.  Note that the final interaction we obtain will be a logical
operator tensored with a complicated interaction on the stabilizer spaces of the concatenated code, and, if we started with single
qubit couplings to a bath, an interaction on the bath.  However our argument has always bounded these operators to have operator
norm $1$ and thus $x_r$ represents the true strength of these interactions.  Finally note that we also retain terms that act on
the excited states of the stabilizer Hamiltonians at all levels.  However these terms do not couple to the ground state.  Since we
will be interested in the case where the interaction strength to excite to these energy levels is exponentially small, these terms
will never be relevant if we start our system in the encoded grounded state.

Thus we see that if the single qubit coupling strength is below $x_*$, then we obtain a doubly exponential suppression of these
terms in the number of levels of concatenation.  Since $r$ levels of concatenation requires $n^r$qubits, this implies that the
effective coupling strength shrinks exponentially as function of the number of qubits.

Finally we note that the new ground state of the concatenated code stabilizer will be close to the ground state of the original
ground state.  At the lowest level of concatenation, our code states will be off by an amplitude of size $O(({x \over J})^2)$.
But these are correctable errors for the next level of concatenation.  By the time we get to the $r$th level of concatenation we
will have an amplitude which is off by $O(({x \over J} )^{2^r})$.  This will not be correctable, but is exponentially small in the
number of qubits.  There is a corresponding threshold associated with this calculation depending on the constant with the
uncorrectable component of the amplitude.

We now point out a few subtleties and drawbacks of our approach.  The first important point is to realize that it is because we are working a regime where the perturbation series converges that we can move from from level to level in our concatenation scheme.   If we were to begin to analyze the effect of the perturbation on the entire Hamiltonian on $n$ qubits for a quantum concatenated code Hamiltonian, it would not be the case that the perturbation theory would converge below some fixed value of the perturbation strength.  This is because the requirement that $x||V||$ is smaller than half the eignevalue gap would require that $x \ll O(1/n)$.  In other words the perturbation would be guaranteed to converge only if the perturbing strength shrunk as a function of the system size.  Note that this does not imply that such an analysis could not be made to work: it is only a statement that the standard approach to perturbation theory would lead to a lack of guaranteed conversion.  We get around this problem by analyzing the effect of the perturbation on each level of the concatenation and showing how an interaction at one level can be reduced at the next level.

A further more important point is that our analysis shows that the perturbation shrinks as $O(2^r)$ where $r$ is the number of levels of concatenation, independent of the code being used.  This seems to defy our intuition that instead of $O(2^r)$ the effect of the perturbation on a $[[n,k,d]]$ code should shrink as $O(d^r)$.   Note, however, that our analysis, is an upper bound on the rate the perturbation shrinks.  We obtain $O(2^r)$ in large part because we need to analyze the effect of the perturbation with respect to the logical qubits used at the next level of concatenation.  In other words, while for a $[[n,k,d]]$ the energy splittings to the ground state occur at order $d$, the fact that there are new eigenstates for the perturbed Hamiltonian means that with respect to the unperturbed eigenstates there are nontrivial interactions.  Whether this is a drawback of our approach or a fundamental obstacle is a question for further investigation.

In review we have showed that single qubit perturbations on a quantum concatenated code Hamiltonian are exponentially suppressed as a function of the number of qubits used in the Hamiltonian for single qubit perturbations which are weaker than some threshold value.  While we have focused on single qubit perturbations, results for two qubit perturbations could also be achieved by using a quantum error correcting code designed to correct two qubit errors.  For the five qubit code, our analysis in Section~\ref{sec:5}, gives a threshold value of the coupling of $x_* \approx 0.0020901 J$ and for the nine qubit subsystem code, $x_* \approx 0.0019936 J$.

\section{Implementing a Concatenated Hamiltonian} \label{sec:gadget}

Quantum concatenated code Hamiltonians on systems of size $n$ require $O(n)$-local interactions.  Since such interactions are not physically reasonable, we must engineer an effective Hamiltonian that yields these interaction terms using terms that are
physically reasonable, like, for instance two qubit interactions.  There are considerable constraints on this procedure imposed,
for example, by the results of~\cite{Haselgrove:03a,Bullock:08a,vanDerNest:08a}.  However the bounds in
~\cite{Haselgrove:03a,Bullock:08a,vanDerNest:08a} are not robust under procedures which use ancillas to engineer many-qubit
interactions using one and two qubit interactions, nor are they robust to interactions which are time dependent.  Here we focus on the later weakness and show how methods with strong pulsed interactions can produce effective evolutions of the desired form~\cite{Zanardi:99b,Viola:99a,Viola:99b,Lidar:07a,Dur:07a}.  This method is very simplistic and has many draw backs.  We include it mostly to show that there is at least some physical possibility of implementing quantum concatenated code Hamiltonians.  

\subsection{Bang-bang techniques for many-qubit simulation of stabilizer Hamiltonians}

\begin{figure}[h]
$$\Qcircuit @C=1em @R=.7em 
{ 
 & \targ & \qw & \qw & \qw & \qw & \qw & \targ & \qw  \\
 & \qw & \targ & \qw & \qw & \qw & \targ & \qw & \qw \\
 & \qw & \qw & \targ & \qw & \targ & \qw & \qw & \qw \\
 & \ctrl{-3} & \ctrl{-2} & \ctrl{-1} & \gate{e^{-iXt}} &\ctrl{-1} & \ctrl{-2} & \ctrl{-3} & \qw
}$$
\caption{A circuit which is equivalent to the unitary $U=\exp(-it X \otimes X \otimes X \otimes X)$.} \label{fig:circuit1}
\end{figure}
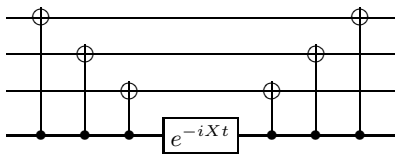

The basic idea is to use a common technique for engineering many-qubit interaction using strong pulses~\cite{Nielsen:00a,Lidar:01b,Lidar:01c}.  Consider the quantum circuit given in FIG.\ref{fig:circuit1}.  This circuit uses two qubit gates conjugated about a single qubit time evolution to produce an effective interaction of a four qubit interaction due to the Hamiltonian $H_{eff}=X \otimes X \otimes X \otimes X$.  One thing to note about this circuit is that, while we have drawn this as a depth seven circuit, the controlled-nots before and after the single qubit gate commute among each other and therefore can be implemented simultaneously.  In other words physically we can implement this circuit using a circuit of depth of three.
A generalization of this procedure~\cite{Lidar:01b,Lidar:01c} is easily seen to allow one to implement any Hamiltonian which has a single term corresponding to an element of the Pauli group.  Depending on how one counts implementation in different single-qubit basis terms, a evolution of the form $\exp(-iPt)$ where $P$ is a $n$ qubit interaction from the Pauli group can be done with a circuit of depth three or depth five.  Note that this constant depth is independent of the weight of the Pauli group member being implemented.  

Having shown how to obtain an effective interaction of the form $\exp(-iPt)$ for $P$ a Pauli group element we now turn to how to construct a concatenated stabilizer code Hamiltonian.   Note that if we concatenate $r$ times, we have $n^r$ qubits and the concatenated stabilizer Hamiltonian is the sum of $n^r-1$.  Stabilizer group elements which all commute with each other.  Recall that for commuting Hamiltonians, $[H_1,H_2]=0$, successive evolutions sum: $\exp(-iH_1t)\exp(-iH_2t)=\exp(-i(H_1+H_2)t)$.  Further note that the concatenated nature of the concatenated stabilizer Hamiltonian can be implemented in a highly parallel nature.  All of the terms at a given level of the concatenation can be implemented in parallel.  This implies that using a circuit of depth $C(n-1)r$ where $C$ is $3$ or $5$ can be used to implement $\exp(-iH_{concat} t)$.  

By repeated application of $U=\exp(-iH_{concat} \Delta t)$ we can now produce a system which evolves according to the evolution operator $U(t)=\exp(-i H_{concat}t )$ in a stepwise fashion in times of step $\Delta t$.  Supposing that our system-bath interaction which we are attempting to suppress has a time scale associated with it of $t_{SB}$, then we need to implement a single step in this process on a timescale much faster than $t_{SB}$ in order for the environment to see an effective evolution on the time scale of $t_{SB}$.  We note that because the depth of our circuit is logarithmic in the number of qubits being used, we can expect that for a given strength of system interactions implementing gates we can obtain a considerable      

The above method for simulating evolution due to a concatenated stabilizer code Hamiltonian is surely naive.  It leaves out detailed consideration of how errors in implementing the simulation are dealt with, and a detailed understanding of whether the assumption of being able to implement bang-bang operations is valid.  This is a subject for future research on quantum concatenated code Hamiltonians.

\section{Computing on Quantum Concatenated Code Hamiltonians and Self-Correction} \label{sec:compute}

In this final section we briefly discuss some issues with regards to performing quantum computation on quantum concatenated code Hamiltonians.

One of the main reasons for studying quantum concatenated code Hamiltonians is, that in addition to providing a ground state
energy manifold which can protect against few qubit perturbations on the ground state, the energy landscape of these systems
contains more than just a constant energy gap to the first excited energy state.  In Kitaev's toric codes in two
dimensions~\cite{Kitaev:97c}, for example, the ground state is robust to perturbation and there is a constant energy gap to the
first excited state, but it is possible, by a series of local operations to cause information on the ground state to be erred
using only a constant amount of energy.  In other words, one can excite the system out of its ground state at a constant energy,
then, at a cost of no energy move across energy levels of the same energy and finally relax back down to the ground state while
causing an error on the information encoded in the ground state.  While this does not have a strong effect on the use of such
systems as a quantum memory, it has been argued that for performing computation on the information encoded in the ground state, such an energy landscape will result in a disordering of quantum computation while the computation is being enacted (see
~\cite{Bacon:06a}).  In contrast, the four spatial dimension toric code has an energy landscape for which an amount of energy
proportional to the energy gap is not sufficient to cause information encoded into the ground state to be
destroyed~\cite{Dennis:02a}.  The energy landscape in this later case is {\em
self-correcting}~\cite{Barnes:00a,Dennis:02a,Bacon:06a}.

Here we note that quantum concatenated code Hamiltonians have an energy landscape which is  self-correcting.  For
example, for the five qubit concatenated code Hamiltonian discussed in prior section, one can see that the lowest weight error
that costs the energy gap $J$ in energy to excite out of the ground state is of weight $2$.  Such an error acts on a single lowest
block of the code (note that a single qubit error on the ground state for the Hamiltonians we have constructed costs $rJ$ energy
since it violates all levels of the concatenation.)  Having caused an error of weight $2$ at a cost of energy $J$, however, an
additional two qubit error which brings the system closer to causing an error which acts non-trivially on the code costs and
energy at least $J$ more (again single qubit errors cost even more.)  In other words, if we imagine applying small weight errors
to the system prepared in its ground state, then an energy barrier of height $rJ$ must be cross before the quantum information
encoded into the ground state can be acted non-trivially upon ($r$ is the number of levels of concatenation in the code.)

Let us make this statement more rigorous.  Suppose that we have a quantum concatenated stabilizer code Hamiltonian using a $[[n,1,d]]$ code, and this code has been concatenated $r$ times.  Consider the set of correctable errors on this concatenated code which are corrected using the block nature of the code (that is the lowest level of concatenation is corrected first, and then the next level is corrected, etc.)  Consider such a correctable error, $E$, and denote $c$ as the number of times in the concatenated code correction where a non-trivial error correction procedure is applied and succeeds in correcting the relevant block (disregard corrections which fail or are trivial.)  Then $E$ acting on the ground state of the code must go to a state of energy at least $wJ$.  This is because each correctable error which is correctly fixed must be able to flip at least one stabilizer generator for the relevant block being error corrected, and this must correspond to a minimal change of $J$ for the concatenated quantum code Hamiltonian.  Thus we know that if a code is concatenated $r$ times, there is an energy barrier of height $rJ$ which must be overcome before an uncorrectable error is reached. 

It is interesting to compare the energy barrier for the concatenated stabilizer code Hamiltonian's with the case of the four-dimensional toric code\cite{Dennis:02a}.  Here the height of the barrier scales as $O(\log N)$ for $N$ qubits while for the toric code in four-dimensions the barrier of the height scales as $O(N^{1/4})$ for $N$ qubits.  Whether the slower scaling for the concatenated code case leads to less robust behavior is an interesting open question.

Thus we can imagine how to perform at least some of the basic computations on the quantum concatenated code Hamiltonians on the ground state.  For example, suppose that we wish to enact an encoded Pauli operator on the ground state.  Such operators for
stabilizer codes are members of the Pauli group.  Thus one way to apply this operator is to locally apply the relevant Pauli
operator to the desired qubits.  If one does this in a bang-bang fashion~\cite{Viola:98a}, where the strength of this application
is much stronger than the natural energy of the quantum concatenated code Hamiltonian ($O(J)$), then one can enact the relevant
Pauli group operator on the ground state.  However, performing bang-bang pulses will inevitably lead to errors during the creation
of the local Pauli group operator.  For example, timing errors may result in over or under rotations of the qubits.  For a similar
procedure with a simple gapped system like Kitaev's two dimensional toric code, such errors will quickly destroy the quantum
information in the ground state.  However, for self-correcting systems, these errors will cause the system to be excited out of
the ground state, but if the errors are few enough in nature, then the system can quickly relax back toward the ground state and restore the quantum information.  Note that it is likely that one would still require that the system remain at temperature below the
energy gap to the first excited state, since occupation of the higher energy levels can cause subsequent quantum operations to
fail.

\section{Conclusion and Future Directions} \label{sec:conc}

We have introduced a new class of physical systems that provide protection for quantum information encoded into the energy levels of these systems called quantum concatenated code Hamiltonians.  We have rigorously shown that these Hamiltonians offer protection against the effects of perturbing interactions, showing a increase in protection which scales exponentially in the number of qubits of the system.  Further the energy landscape of these Hamiltonians differs from topological quantum systems, which exhibit a energy gap to the first excited states, but can be disordered with only a constant amount of further energy once such excited states are created.  Quantum concatenated code Hamiltonians represent a new method for physically protecting quantum information which is closely connected to the traditional method of using concatenation to achieve fault-tolerant quantum computing.  This close connection should allow for many of the tools and techniques of this standard method to be applied to physical methods for building a robust quantum system.

There is much work, however, that needs to be performed before the ideas described in this paper become a viable path toward
constructing a large scale robust quantum computer.  The first, and most pressing issue, is that while we have used one and two
qubit interactions gates to construct a bang-bang simulation of the Hamiltonian, it would be better if we could find a way to more directly implement these Hamiltonians.  One method which is promising is to leverage existing results about perturbation theory gadgets~\cite{Kempe:06a,Oliveira:05a,Jordan:07a}.  However, to date, perturbation theory gadgets cannot engineer interactions of great enough simultaneous strength and size to realize concatenated code Hamiltonians.  

A second open question is how to perform universal quantum computation on quantum concatenated code Hamiltonians.  There are two issues here.  One is when the gates to be enacted are performed in a pulsed fashion.  While we have discussed the enaction of Pauli group operations, a larger set of gates is required in order to achieve universal quantum computation.  A related question is  how the energy landscape of a quantum concatenated code Hamiltonian can be used to suppress computational errors when enacted pulsed gates on the quantum information. A detailed understanding of the statistical physics of quantum concatenated code Hamiltonian would be helpful for this endeavor~\cite{Alicki:06a}.  A second approach to performing computation using these Hamiltonians is to use them as building blocks for the recently discovered methods for universal adiabatic quantum computation~\cite{Aharonov:04a,Kempe:06a,Mizel:06a,Mizel:01a,Mizel:02a}.  Such adiabatic evolutions may be more suitable for a paradigm where an always on Hamiltonian is used to protect quantum information.   

Finally we note that while quantum concatenated code Hamiltonians have Hamiltonians of a highly nonlocal nature, there is reason to believe that local few qubit Hamiltonians can give rise to these Hamiltonians.  In particular we note that concatenated code states fit naturally within the ansatz states of  entanglement renormalization~\cite{Vidal:05a}.  Thus concatenated code states serve as candidates, in addition to toric code states~\cite{Vidal:08a}, for fixed points in this renormalization method.  

\acknowledgements

This work was funded by NSF grant number 0523359.  Dave Bacon is also supported by ARO/NSA quantum algorithms grant number W911NSF-06-1-0379 and NSF grant number 0621621.  Useful conversations with Kenneth Brown, Stephen Jordan, David Poulin, and John Preskill are gratefully acknowledged.  We thank Stephen Jordan and Ed Farhi for sharing an early version of~\cite{Jordan:07a}, and Jonathan Vos Post for catching a numeric error in a previous draft of the paper.

\appendix

\section{Kato's Perturbation Theory} \label{sec:kato}

Kato~\cite{Kato:49a} and Bloch~\cite{Bloch:58a} long ago developed a rigorous perturbation theory that allows one to derive
effective Hamiltonians for perturbing interactions that is particular useful when the spectrum of the unperturbed Hamiltonian has
widely separated eigenvalues (wide in comparison roughly to the strength of the perturbation.)   The reader is referred to Kato's
work~\cite{Kato:49a} for details of the former perturbation theory and to Bloch, and Jordan and Farhi~\cite{Bloch:58a,Jordan:07a}
for details of the later. Here, for completeness, we list the main results of Kato's perturbation theory and in particular use
the bounds derived in an early version of~\cite{Jordan:07a} for errors in truncating this perturbation theory.

Let $H_0$ be a Hamiltonian with an eigenvalue decomposition $H_0=\sum_i E_i \Pi_i$ where $E_i$ the energy of the $i$th eigenspace
and $\Pi_i$ is a projector onto this $i$th, possibly degenerate, eigenspace.  We assume that the $\Pi_i$ form a complete set of
projectors onto the Hilbert space we are working on, $\Pi_i \Pi_j=\delta_{ij} \Pi_i$ and $\sum_i \Pi_i=I$.  We will consider a
perturbation of $H_0$ by $xV$, where $x$ is the strength of the perturbation, so that the full Hamiltonian is $H_x=H_0+xV$.  We
will be concerned with the situation where $H_0$ has discrete energy levels that are well separated from each other, but possibly
degenerate.  In this case as a function of $x$, the eigenvalues of $H$ split away from their values at $H_0$ in a well behaved
manner.  If we focus on a particular eigenvalue $\lambda_k$ of $H_0$, then it is possible to obtain an expression for the
Hamiltonian $H_x$ restricted to the eigenvalues arising from the shift/splitting of this $k$th eigenvalue of $H_0$.  If the entire
spectrum of $H_0$ consists of well separated eigenvalues (the exact nature of this condition on the spectrum will be described
below), then we can obtain an expression for the entire effective Hamiltonian.

Begin by defining
\begin{equation}
G_i^{(k)} = \left\{ \begin{array}{ll}
\sum_{j \neq i} {\Pi_j \over (\lambda_j-\lambda_i)^k} & {\rm if~}k>0 \\
-\Pi_i  & {\rm if}~k=0
\end{array} \right. .
\end{equation}
Then the effective Hamiltonian coming from the splitting of the $i$th eigenvalue(s) of $H_0$ is
\begin{equation}
H^{\rm eff}_i=H_x P_i(x) = \lambda_i P_i(x) + \sum_{m=1}^\infty x^m A_i^{(m)}, \label{eq:kato}
\end{equation}
where
\begin{equation}
A_i^{(m)} = (-1)^{m-1} \sum_{(m-1)} G^{(k_1)}_i V G^{(k_2)}_i V \cdots V G^{(k_{m+1})}_i.  \label{eq:term}
\end{equation}
and $P_i(x)$ is the projector onto the eigenvalues of $H_x$ arising from the $i$th eigenvalue(s) of $H_0$.  The notation
$\sum_{(m-1)}$ means that the sum should be taken over all sets of integers $k_j \geq 0$ such that $\sum_{i=1}^{m+1} k_i = m-1$.
In our calculations we will often use the first few terms the sum in Eq.~(\ref{eq:kato}).   We list these here for ease of
reference,
\begin{eqnarray}
A_i^{(1)} &=& G_i^{(0)} V G_i^{(0)}, \nonumber \\
A_i^{(2)} &=& -G_i^{(0)} V G_i^{(1)} V G_i^{(0)} - G_i^{(1)} V G_{i}^0 V G_i^{(0)} - G_i^{(0)} V G_i^{(0)} V G_i^{(1)}, \nonumber \\
\end{eqnarray}
Further the projector onto the eigenvalues of $H_x$ arising from the $i$th eigenvalues(s) of $H_0$ has a perturbative expansion
\begin{equation}
P_i(x)=\Pi_i+\sum_{n=1}^\infty B_i^{(n)} x^n, \label{eq:kato2}
\end{equation}
where
\begin{equation}
B_i^{(m)} = (-1)^{m-1} \sum_{(m)} G^{(k_1)}_i V G^{(k_2)}_i V \cdots V G^{(k_{m+1})}_i.  \label{eq:projterm}
\end{equation}
Eqs.~(\ref{eq:kato}) and (\ref{eq:kato2}) are the main expression of importance in this appendix.  When $x||V||$ is less than half
the distance from $\lambda_i$ to its nearest other eigenvalue of $H_0$, $\lambda_j$, this sum is guaranteed to converge and is
correctly reproduces the spectrum of $H_x$ arising from the splitting/shifting of the $i$th eigenvalue of $H_0$.  Sometimes this sum is exactly evaluatable and thus one can obtain a closed form expression for $H_x$.  On the other hand, the
series is more widely used to obtain a perturbative expansion in $x$, truncating the sum in Eq.~(\ref{eq:kato}) at a fixed $m$. In
this case it is important to know the magnitude, and often the form, of the remaining terms in this sum.  By an analysis perform
in an earlier draft of Jordan and Farhi~\cite{Jordan:07a} one can show that if the eigenvalue $i$ is separated from all of other
eigenvalues by at least $\Delta$, then if we truncate the series at the $p$th term, then the remaining terms have an operator norm
of no greater than
\begin{equation}
\left \|\sum_{m=p+1}^\infty x^m A_i{(m)} \right \| \leq {\Delta \over 2} { \left({4 x ||V|| \over \Delta}\right)^{p+1} \over 1-{4
x ||V|| \over \Delta}},
\end{equation}
which, for $x ||V|| \leq {\Delta \over 8}$ is bounded by
\begin{equation}
\left \|\sum_{m=p+1}^\infty x^m A_i{(m)} \right \| \leq \Delta \left({4 x ||V|| \over \Delta}\right)^{p+1}.
\end{equation}
Similarly for the projector onto the $i$ energy eigenspace we can bound the error as
\begin{equation}
\left \|\sum_{m=p+1}^\infty x^m B _i{(m)} \right \| \leq 2 \left({4 x ||V|| \over \Delta}\right)^{p+1}.
\end{equation}
We note that these bounds are not quite tight, but are reasonable for our purposes.  

\bibliographystyle{hunsrt}
\bibliography{../../bigref/newbigref}

\end{document}